\documentclass[sigplan,10pt]{acmart}

\makeatletter
\def\mdseries@tt{m}
\makeatother

\acmYear{2018}\copyrightyear{2018}
\setcopyright{acmlicensed}
\acmConference[PPoPP '18]{PPoPP '18: Principles and Practice of Parallel Programming}{February 24--28, 2018}{Vienna, Austria}
\acmBooktitle{PPoPP '18: Principles and Practice of Parallel Programming, February 24--28, 2018, Vienna, Austria}
\acmPrice{15.00}
\acmDOI{10.1145/3178487.3178494}
\acmISBN{978-1-4503-4982-6/18/02}

\newif\iffull
\fulltrue
\iffull
\makeatletter
\@ACM@authorversiontrue
\makeatother
\fi

\usepackage{listings}
\usepackage{code}

\usepackage{booktabs}

\newcommand{\secref}[1]{Section~\ref{sec:#1}}
\newcommand{\figref}[1]{Figure~\ref{fig:#1}}
\newcommand{\figreftwo}[2]{Figures \ref{fig:#1} and~\ref{fig:#2}}

\newcounter{remark}[section]
\newenvironment{nop}{}{}

\newcommand{\la}{\ensuremath{\leftarrow}}
\newcommand{\cd}[1]{{\lstinline!#1!}}

\lstdefinelanguage{pseudocode}
{
  morekeywords={
    type,
    function,
    if,
    else,
    while,
    for,
    break,
    run,
    in,
    wait,
    return,
    from,
    up,
    down,
    to,
    below,
    and,
    not,
    included,
    excluded,
    assert,
    case,
    of,
    current,
    roots
  },
  xleftmargin=0.75em,
  numbers=left, 
  numbersep=0.5em,
  sensitive=true, 
  morecomment=[s]{(*}{*)}, 
  morestring=[b]" 
}

\begin{document}

\title{Hierarchical Memory Management for Mutable State}
\iffull
\subtitle{Extended Technical Appendix}
\fi

\author{Adrien Guatto}
\affiliation{
  \institution{Carnegie Mellon University}
}
\email{adrien@guatto.org}
\author{Sam Westrick}
\affiliation{
  \institution{Carnegie Mellon University}
}
\email{swestric@cs.cmu.edu}
\author{Ram Raghunathan}
\affiliation{
  \institution{Carnegie Mellon University}
}
\email{ram.r@cs.cmu.edu}
\author{Umut Acar}
\affiliation{
  \institution{Carnegie Mellon University}
}
\email{umut@cs.cmu.edu}
\author{Matthew Fluet}
\affiliation{
  \institution{Rochester Institute of Technology}
}
\email{mtf@cs.rit.edu}



\begin{abstract}
  It is well known that modern functional programming languages are
  naturally amenable to parallel programming.
  Achieving efficient parallelism using functional languages,
  however, remains difficult.
  Perhaps the most important reason for this is their lack of support
  for efficient in-place updates, i.e., mutation,
  which is important for the implementation of
  both parallel algorithms 
  and the run-time system services (e.g., schedulers and synchronization primitives)
  used to execute them.

  In this paper, we propose techniques for efficient mutation in
  parallel functional languages.
  To this end, we couple the memory manager with the thread scheduler to make reading and updating data allocated by nested threads efficient.
  We describe the key algorithms behind our technique, implement them
  in the MLton Standard~ML compiler, and present an
  empirical evaluation.
  Our experiments show that the approach performs well, significantly
  improving efficiency over existing functional language implementations.

\end{abstract}

\sloppy
\maketitle
\renewcommand{\shortauthors}{A.~Guatto, S.~Westrick, R.~Raghunathan, U.~Acar, and M.~Fluet}


\section{Introduction}
\label{sec:introduction}

With the proliferation of parallel hardware, functional programming
languages, such as Haskell and the ML family (OCaml, Standard~ML),
have received much attention from academia and industry.
Even non-functional languages today such as C++, Python, and Swift
support certain features of functional languages, including
higher-order functions and, sometimes, rich type systems.
An important virtue of strongly typed functional languages is
their ability to distinguish between pure and impure code.
This aids in writing correct parallel programs by making it easier to
avoid race conditions, which can become a formidable challenge in
languages whose type systems don't distinguish between mutable and
immutable data.

In the sequential realm, functional languages compete well with
other garbage-collected languages such as Java and Go,
often running within a factor of 2 or 3 and sometimes even faster.
In many cases, functional languages even compete well with the C
family, where memory is managed  by the
programmer~\cite{wadler-why-1998}.

In the parallel realm, however, the gap between functional and
imperative languages is significantly larger.
One reason for this is the (poor) support for mutation in
parallel functional languages.
A reality of modern hardware is that imperative algorithms can
perform significantly better than pure functional algorithms by using
constant-time random accesses and updates.
Even when a parallel algorithm has a pure functional interface
(immutable inputs and immutable outputs), it can be more efficient to
use mutation internally.
For example, the efficiency of a pure functional parallel merge-sort
can be significantly improved by reverting to a sequential imperative
quick-sort for small inputs.
Committing to pure functional algorithms only does not completely
avoid mutation: a language run-time system uses mutation to implement
crucial facilities such as (thread) schedulers and synchronization
primitives, which require communication between processors via shared
memory.

Even though mutation is crucial for efficiency, it remains poorly
supported in parallel functional languages and remains as an active
area of research~(see \secref{related}).
For example, the Manticore project has developed rich extensions to
the ML language to support parallelism
but has focused on purely functional code where the programmer cannot
use mutation~\cite{manticore-implicit-11,abfr-manticore-gc-11}.
Other ML dialects such as OCaml and SML\# continue to remain
primarily sequential languages, though there is ongoing work in
extending them to support modern parallelism features.
Like Manticore, Haskell has a relatively rich set of parallelism
features and its runtime must support efficient
mutation~\cite{marl11}.
Writing efficient parallel programs in Haskell, however, remains difficult
in part because of lazy evaluation~\cite{newton-ryan-2017}.

This state of the art raises the question of whether
functional programming can be extended to support mutable data and
parallelism efficiently.
At the highest level of abstraction, this is a challenging problem
because its parts---parallelism and efficient memory management---are
individually challenging.
The problem is further complicated by the fact that functional
languages allocate and reclaim memory at a much faster
rate than other
languages~\cite{appe89,doli93,doli94,gonc95,gonc95a,appe96,abfr-manticore-gc-11,marl11}.

In this paper, we propose techniques for supporting mutable data
efficiently within the run-time system of nested-parallel functional
languages, focusing on strict languages in the style of the ML family.
Our approach builds upon that of \emph{hierarchical heaps}, a memory management
technique developed in prior work~\cite{hiheap-2016}.
The basic idea is to organize memory so that it mirrors the structure of the parallel
computation.
Specifically, each thread is assigned its own
heap in a hierarchy (tree) of heaps which grows and shrinks as threads are
forked and joined.
Threads allocate data in their own heaps, and can read and update objects in
ancestor heaps (including their own).
A key invariant is that data in non-ancestor heaps remains unreachable
to a thread.
To enforce this invariant, we propose a \emph{promotion} technique for copying
data upwards in the hierarchy as necessary.

Our approach has several important benefits.
First, threads can allocate, read, and update mutable objects
in their heap, without synchronization or copying.
This allows local mutable objects to be used efficiently.
Second, because
heaps are associated with threads rather than
processors, a thread can be migrated between processors without
copying data.
These two properties contrast with the predominant approach to
memory management in parallel functional languages with local heaps, where both mutation and
thread migration require copying~\cite{doli93,doli94,abfr-manticore-gc-11,marl11}.
Third, our techniques introduce no overhead for reads of immutable objects,
which are pervasive in functional languages.
Finally, any thread can collect its heap independently and, more
generally, any subtree of heaps in the hierarchy could
be collected independently.

The contributions of this paper include techniques and algorithms for
handling mutable data in hierarchichal heaps (\secref{algo}), an implementation
extending the MLton whole-program optimizing compiler for Standard
ML~\cite{mlton}, and an empirical evaluation considering a number of
both pure and imperative benchmarks (\secref{exp}).
Our results show that these techniques can be implemented efficiently and
can perform well in practice.

\iffull
\else
Due to space restrictions, we omit extraneous details, providing them
in a technical appendix~\cite{this-paper-online}.
\fi

\section{Overview}
\label{sec:background}

We present a brief overview of our techniques, using a simple example to
illustrate both the programming model and details of memory management.
In the process, we introduce terminology that will be used
throughout the paper.

\begin{figure}
\begin{minipage}[b]{0.45\textwidth}
\begin{lstlisting}[morekeywords=par]
val GRAIN = ...
fun inplaceQSort s =  ... 
fun msort s =
  if Seq.length s <= GRAIN
    then let val a = Seq.toArray s
             val () = inplaceQSort a
         in Seq.fromArray a end 
  else let val (l, r) = Seq.splitMid s
           val (l', r') = par (msort l, msort r)
       in Seq.merge (l', r') end
\end{lstlisting}
\caption{Code for parallel  imperative merge sort.}
\label{fig:msort-imperative}
\end{minipage}

\medskip

\begin{minipage}[b]{0.45\textwidth}
\vspace{0in}
\centering
\includegraphics[width=0.95\textwidth]{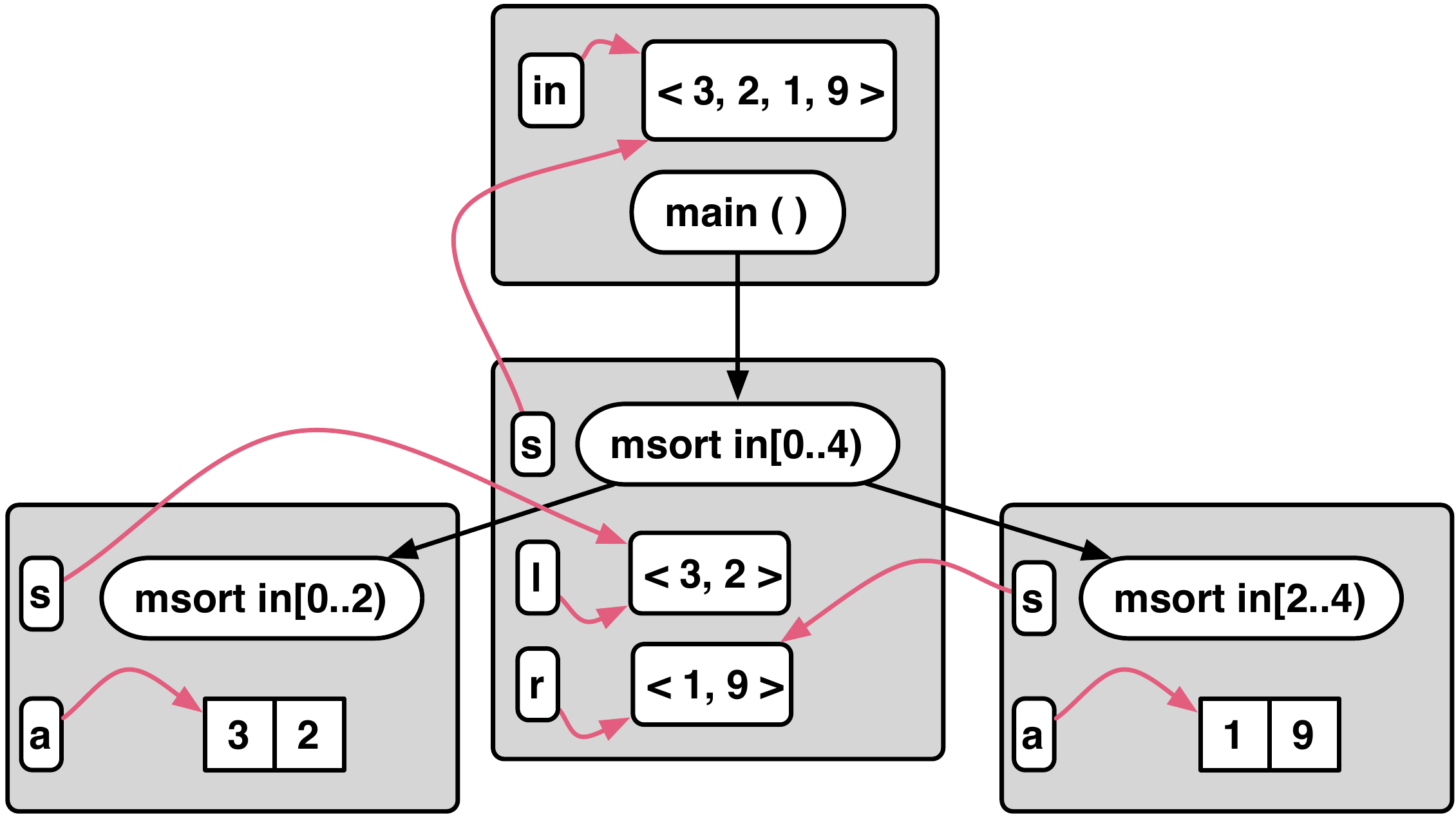}
\caption{Hierarchical heap example for \cd{msort}.}
\label{fig:msort-imperative-hh}
\end{minipage}
\end{figure}



Consider the parallel merge sort in \figref{msort-imperative}.
The implementation uses an immutable sequence data structure
provided by a module \cd{Seq}, whose details we omit.
To sort an input sequence, the function \cd{msort} first checks its
length.
If the length is less than some constant \cd{GRAIN}, then the function uses an
imperative in-place sequential quicksort to sort the input.
Otherwise, the input is split into
two halves and two recursive calls are performed in parallel.

Parallelism is exposed by the programmer through the
\lstinline[morekeywords=par]{par} construct, which creates new \emph{tasks}.
Initially, there is only one task (corresponding to the execution of the entire
program); thus all user code implicitly runs within the context of some task.
With \cd{par}, a task may spawn two new tasks.
This establishes a parent/child nesting relationship where a parent task must
wait until both its children complete before continuing its execution.
Tasks are managed by a \emph{scheduler}, which strives to minimize the
completion time of the program by mapping tasks to processors.

As in
Raghunathan et al.~\cite{hiheap-2016},
to support parallel automatic memory management, each task is assigned its own
\emph{heap} in which it allocates new data.
Heaps are organized into a \emph{hierarchy} (tree) with the same
parent/child relationships as their associated tasks.
When both children of a task complete, their heaps are logically merged with
the parent heap, allowing the parent to continue with child data stored locally
as though the children had never existed.

\figref{msort-imperative-hh} illustrates an example where \cd{msort} is called
with the input sequence $\langle 3, 2, 1, 9 \rangle$ and \cd{GRAIN} = 2.
Tasks are drawn as ellipses connected by straight, black arrows pointing from
parent to child.
The grey boxes drawn around each task delimit its heap, and red curved arrows
show pointers in memory.
At the root of the hierarchy is the initial task \cd{main} which allocated the
input sequence and then called \cd{msort}.
The middle task corresponds to this initial call of \cd{msort}, which split
the input into two sequences \cd{l} and \cd{r}, allocated locally.
The two leaf tasks are the parallel recursive calls of \cd{msort}, which
have copied their inputs to local arrays \cd{a}.

Consider the following definition.
We say that the hierarchy is \emph{disentangled} if,
for any pointer from an object $x$ in heap $h_x$ to another object $y$ in heap
$h_y$, $h_y$ is either equal to or an ancestor of $h_x$.
In other words, in order to be disentangled, the hierarchy must not contain
\emph{down-pointers} that point from ancestor to descendant, nor may it contain
\emph{cross-pointers} that point between unrelated heaps (such as between
two siblings).
In a disentangled hierarchy, the lack of pointers into leaf heaps
allows us to garbage collect the leaves independently, without synchronizing
with other tasks, and in parallel with other leaf-heap collections.
More generally, any two disjoint sub-trees of heaps may be collected
independently in parallel (although the tasks within a sub-tree must
cooperate).
The execution of \cd{msort} in~\figref{msort-imperative-hh} is
disentangled, thus the two leaves of the hierarchy could both be independently
collected in parallel;
for example, collecting the array \cd{a} after computing
\cd{Seq.fromArray a} but before joining with the parent task.
It can be shown that all purely functional programs naturally exhibit
disentanglement~\cite{hiheap-2016}.

In the presence of arbitrary mutation, disentanglement is not guaranteed.
Consider for example a mutable reference \cd{r} which is allocated at a
parent task and then passed to two child tasks.
One child could update \cd{r} to point to a locally allocated object,
creating a down-pointer.
The second child could then read from \cd{r} to create a cross-pointer.

To enforce disentanglement, we propose a \emph{promotion} technique.
The basic idea behind promotion is to detect when a down-pointer would be
created in the hierarchy, and first promote (copy) the lower object upwards so
that it lies in the same heap as the mutable object.

There are a number of challenges associated with promotion.
In particular, promotion duplicates objects, which complicates the identity of
mutable objects.
We solve this issue by distinguishing one copy as the
\emph{master copy}, to which all accesses are redirected through
\emph{forwarding pointers}.
Additionally, when data being promoted contains pointers to other objects,
all transitively reachable data might need to be promoted.
This introduces concurrency into the system even when none exists in the user
code, since a task might access an object which is in scope of
an in-progress promotion.

In our implementation, we prioritize the efficiency of updates to local
objects.
This facilitates an important idiom of practical parallelism where the
overhead of parallelism is amortized by switching to a fast sequential algorithm
on small inputs.
As the fastest sequential algorithms on modern hardware are often imperative,
local updates are thus crucial for efficiency.
The \cd{msort} program exemplifies this idiom, utilizing a fast imperative
quicksort on small inputs.
Indeed, in our results (\secref{results}), we see that \cd{msort}
can be up to twice as fast as a purely functional alternative.

\section{Hierarchical Heaps for Mutable Data}
\label{sec:algo}
\label{sec:alg}

\lstset{language=pseudocode}

\subsection{Programming Model}

Our proposal deals with the full ML language, including mutable data, extended
with nested task parallelism.
We extended an existing ML compiler to reduce this programming model to a small
set of high-level operations that we implemented in the runtime system.
In what follows, we describe the interface of these high-level operations in
abstract terms and explain the challenges we face when implementing them
in the context of hierarchical heaps.

\paragraph{High-Level Types.}

\begin{figure}\lstset{language=pseudocode}
\begin{lstlisting}
type task
type data
type objptr
type field
type thunk = unit -> objptr
function forkjoin: thunk * thunk -> objptr * objptr
function alloc: field list -> objptr
function readImmutable: objptr * field -> data
function readMutable: objptr * field -> data
function writeNonptr: objptr * field * data -> unit
function writePtr: objptr * field * objptr -> unit
\end{lstlisting}
\caption{High-Level Operations.}
\label{fig:high-level-operations}
\end{figure}

\figref{high-level-operations} describes the types and operations we use to
implement nested-parallel ML programs.
While such operations are typically programmed in C or some other low-level
language, here for the sake of simplicity we specify them in ML-like pseudocode.


Tasks compute on~\emph{data}.
The exact definition of data does not matter at this level of
description; we can assume that it consists in machine integers, floating-point
numbers, pointers, etc.
Amongst general data, we distinguish the type~\texttt{objptr} of~\emph{object
  pointers}, that is, of pointers to allocated objects, or simply~\emph{objects}.
Objects correspond to ML data types allocated during the course of execution,
such as cons cells.
For our purposes, an object consists of a finite list of~\emph{fields} storing
its content.
Objects are used for communication between tasks.
The body of a task is a~\emph{thunk}, that is, a function expecting no argument
and returning its result as an object pointer.

\paragraph{High-Level Operations.}

The types described above are used through six high-level operations.
One of them deals with task management, the others with objects.

The compiler elaborates the~\lstinline[morekeywords=par]{par} keyword into
calls to the~\texttt{forkjoin} operation.
This operation takes two thunks, creates one task for each, and runs both tasks
to completion, in parallel.
This establishes a nesting relationship: we say that the task
calling~\texttt{forkjoin} is the~\emph{parent task}, while the tasks created by
the operation are the~\emph{children tasks}.
The parent task is suspended until both children return, at which point it
receives their results and resumes.

Allocation arises from a variety of ML features:~constructors of algebraic data
types, explicit initialization of references and arrays, closures, etc.
The~\texttt{alloc} operation allocates a new object.
The caller describes the list of fields of the new object and receives a
pointer to it as a result.

The remaining primitives deal with reading from and writing to objects.
Since the type system of ML distinguishes between mutable and immutable data, we
have several reading and writing operations, depending on the type of the data
being read or written.
Later in this section, we will exploit these distinctions for efficiency
purposes.

First, we distinguish between reading immutable data with the
\texttt{readImmutable} operation and reading mutable data with the
\texttt{readMutable} operation.
Both operations take a pair of an object pointer and a field descriptor, and
read the data held in the corresponding field of the object.
If the field does not exist, the result is undefined.

Second, when writing to mutable data, we distinguish between writing non-pointer
data~(e.g., machine integers) and writing object pointers.
The former takes the object and field to be written to, as well as the data to
write.
The latter takes the same argument, except that the data to write must be an
object pointer.
Note that, in the type signature of~\texttt{writeNonptr}, we specify the
general~\texttt{data} type, which includes pointers; however, the behavior of
the function is undefined if it is actually passed an object pointer.

\subsection{Low-Level Primitives.}

\begin{figure}\lstset{language=pseudocode}
\begin{lstlisting}
function currentTask: unit -> task
function getField: objptr -> field -> ptr
function fields: objptr -> field list
function ptrFields: objptr -> field list
function nonptrFields: objptr -> field list
function fwdPtr: objptr -> ptr
function hasFwdPtr: objptr -> bool
type heap
function heapOfTask: task -> heap
function newChildHeap: heap -> heap
function joinHeap: heap * heap -> unit
function depth: heap -> int
function freshObj: heap * field list -> objptr
function heapOf: objptr -> heap
function lock: heap -> { READ, WRITE } -> unit
function unlock: heap -> unit
\end{lstlisting}
\caption{Low-level Primitives.}
\label{fig:low-level-interface}
\end{figure}

Our algorithms implement the interface of~\figref{high-level-operations} using
hierarchical heaps.
To do so, they rely on a number of low-level operations
(\figref{low-level-interface}) provided by the runtime system.

\paragraph{Task-Related Operations.}

Calling~\texttt{currentTask} retrieves the currently running task.
Since user code is always implicitly running within some task, this never fails.

\paragraph{Memory-Related Operations.}

The~\texttt{getField} operation returns a pointer to the specified field in an
object pointer.
This pointer does not point to~(the beginning of) an object, but rather
inside it.
For this reason, the result type of~\texttt{getField} is the abstract
type~\texttt{ptr}, rather than~\texttt{objptr}.

Calling~\texttt{fields(o)} operation returns all the fields of the memory
object.
Its variants~\texttt{ptrFields} and~\texttt{nonptrFields} return respectively
the fields which hold object pointers and those which do not.

Finally, each object comes equipped with a special field for storing
a~\emph{forwarding pointer}.
Forwarding pointers are a classic ingredient used in copying
collectors~\cite{jone11} to perform bookkeeping during collection.
An object may or may not have a valid forwarding pointer in its dedicated field.
This can be tested using the~\texttt{hasFwdPtr} operation.
The address of the field can be obtained by calling~\texttt{fwdPtr}.

\paragraph{Heap-Related Operations.}

The first group of operations on heaps~(abstract type~\texttt{heap}) manages
their relationships with tasks as well as with each other.
Calling~\texttt{heapOfTask(t)} returns the heap associated with task~\texttt{t}.
Like tasks, heaps are organized into a hierarchy which grows and shrinks through
the~\texttt{newChildHeap} and~\texttt{joinHeap} primitives.
Calling~\texttt{depth(h)} returns the depth of a heap~\texttt{h} in the
hierarchy:~the root is at depth zero, its children at depth one, etc.

The operation~\texttt{freshObj} allocates a new object in the specified heap and
with the specified fields.
Correspondingly, the heap in which an object was allocated can be retrieved by
the \texttt{heapOf} operation.

Finally, in order to deal with concurrency issues, every heap comes equipped
with a readers-writers lock~\cite{herlihy-shavit-2011}.
Such locks can be held in reading mode by several threads simultaneously, but by
a single thread in writing mode~(excluding any other readers or writers).
The lock associated with a heap~\texttt{h} can be acquired by
calling~\texttt{lock(h,m)}, with~\texttt{m} being the~\texttt{READ}
or~\texttt{WRITE} mode, and then released by calling~\texttt{unlock(h)}.

\subsection{Implementation of the High-Level Primitives}

We can now explain our algorithms as implementations of the high-level
operations in terms of the low-level ones, starting with an overview of the
challenges involved.

\paragraph{Challenges.}

Our goal is to ensure that disentanglement of the hierarchy holds in the
presence of calls to~\texttt{writePtr(obj,field,ptr)}.
If implemented na\"ively as~\texttt{*getField(obj,field)~$\leftarrow$~ptr},
such writes would create down-pointers when~\texttt{heapOf(obj)} is an ancestor
of~\texttt{heapOf(ptr)}.
In order to enforce disentanglement in this situation, our
implementation of~\texttt{writePtr} promotes~(copies) the object
at~\texttt{ptr}, as well as all objects reachable from it,
into~\texttt{heapOf(obj)}.
The address of the copy can then be written into~\texttt{*getField(obj,field)}.

In the presence of repeated writes of~\texttt{ptr} to objects held in heaps of
decreasing depth, the object at~\texttt{ptr} might be promoted several times.
Ultimately, all copies but one should be eliminated, and pointers to them
updated to point to the remaining one.
Thus, we need a way to link an object with its copies.
We use forwarding pointers to do so, and organize all the existing copies of an
object into a singly-linked list.

All the copies of an object are equivalent as far as immutable fields are
concerned, since by definition their content cannot change.
In contrast, reads and writes of mutable fields cannot treat all copies as
equivalent, for this could lead to lost updates~(if an updated copy is then
eliminated) or inconsistent reads~(if a task updates one copy and then reads
another one).
Thus, we have to perform all mutable accesses on a unique, authoritative copy of
an object, that we call its~\emph{master copy}.
Given that all copies are arranged into a linked-list, we choose to take the
last element of that list, that is the one in the shallowest heap, as the master copy.

A further difficulty is that several tasks might be trying to copy the same
data, and thus update the relevant forwarding pointers, simultaneously.
A task might also be updating forwarding pointers while another one is
traversing them to find the master copy.
Our algorithms avoid synchronization issues by acquiring the locks of the heaps
they traverse during mutable accesses, from deepest to shallowest.
Additionally, we propose several fast-paths avoiding locking in common
cases.

\begin{figure}
\begin{lstlisting}[language=pseudocode]
function forkjoin (f, g) =
  heap $\la$ heapOfTask(currentTask ())
  heap_f $\la$ newChildHeap(heap)
  heap_g $\la$ newChildHeap(heap)
  (r_f, r_g) $\la$ run $t \in \{$f$,$g$\}$ in heap$_t$ and wait
  joinHeap(heap, heap_f); joinHeap(heap, heap_g)
  return (r_f, r_g)
\end{lstlisting}
\caption{Fork/join.}
\label{fig:alg-0}
\end{figure}

\begin{figure}
\begin{lstlisting}[language=pseudocode]
function alloc (fields) =
  return freshObj(heapOfTask(currentTask()), fields)
function readImmutable (obj, field) =
  return *getField(obj, field)
function findMaster (obj) =
  while true:
    while hasFwdPtr(obj): obj $\la$ *fwdPtr(obj)
    lock(heapOf(obj), READ)
    if not hasFwdPtr(obj): return obj
    unlock(heap(obj))
function readMutable (obj, field) =
  res $\la$ *getField(obj, field)
  if not hasFwdPtr(obj): return res
  obj $\la$ findMaster(obj)
  res $\la$ *getField(obj, field)
  unlock(heapOf(obj))
  return res
function writeNonptr (obj, field, val) =
  *getField(obj, field) $\la$ val
  if not hasFwdPtr(obj): return
  obj $\la$ findMaster(obj)
  *getField(obj, field) $\la$ val
  unlock(heapOf(obj))
\end{lstlisting}
\caption{Allocation, reads, non-pointer writes.}
\label{fig:alg-1}
\end{figure}

\begin{figure}
\begin{lstlisting}[language=pseudocode]
function writePtr (obj, field, ptr) =
  if heapOf(obj) = heapOfTask(currentTask())
     and not hasFwdPtr(obj):
    *getField(obj, field) $\la$ ptr
    return
  obj $\la$ findMaster(obj)
  if depth(heapOf(obj)) $\ge$ depth(heapOf(ptr)):
    *getField(obj, field) $\la$ ptr
    unlock(heap(obj))
    return
  unlock(heap(obj))
  writePromote(obj, field, ptr)
function writePromote (obj, field, ptr) =
  assert (depth(heapOf(obj)) < depth(heapOf(ptr)))
  prev $\la$ ptr
  lock(heapOf(prev), WRITE)
  while true:
    for h from heapOf(prev) excluded
          up to heapOf(obj) included: lock(h, WRITE)
    if not hasFwdPtr(obj): break
    else:
      prev $\la$ obj
      obj $\la$ *fwdPtr(obj)
  promotedPtr $\la$ promote(heapOf(obj), ptr)
  *getField(obj, field) $\la$ promotedPtr
  for h from heapOf(obj) included
        down to heapOf(ptr) included: unlock(h)
function promote (heap, obj) =
  if depth(heapOf(obj)) $\le$ depth(heap): return obj
  if hasFwdPtr(obj):
    return promote(heap, *fwdPtr(obj))
  newObj $\la$ freshObj(heap, sizeOf(obj))
  *fwdPtr(obj) = newObj
  for field in nonptrFields(obj):
    *getField(newObj, field) $\la$
      *getField(obj, field)
  for field in ptrFields(obj):
    *getField(newObj, field) $\la$
      promote(heap, *getField(obj, field))
  return newObj
\end{lstlisting}
\caption{Pointer writes and promotion.}
\label{fig:alg-2}
\end{figure}

\paragraph{Fork-join.}

\figref{alg-0} shows a na\"ive implementation of the fork/join operation.
First, create a heap attached to the heap of the running task for each child
task.
Then, run in parallel each~$t$ within its heap~\cd{heap$_t$},
for~$t \in \{ \mathtt{f}, \mathtt{g} \}$, and wait for them to complete~(the
exact realization of these operations depending on the scheduler at hand).
Finally, join both child heap with their parent, and pass the results returned
by each task to the caller.
Joining heaps can be done without physically copying data.

\paragraph{Allocation.}

Our implementation of~\texttt{alloc} allocates the new object in the heap of the
currently running task~(l.~2).

\paragraph{Reading Immutable Data.}

ML programs read immutable data when destructuring values such as tuples or
lists, e.g. through projections or pattern matching.
Since these are very common operations, it is important to support them
efficiently.
Fortunately, all the potential copies of the same object hold the same value in
their immutable fields, by definition.
Thus,~\cd{readImmutable(obj,field)} does not care about the forwarding pointer
slot of~\cd{obj} and can always access the contents of~\cd{field} without any
indirection~(l.~4).

\paragraph{Finding Master Copies.}

After several promotions occur, objects may exist in multiple copies linked in a
chain by their forwarding pointers.
This chain has to be taken into account when accessing mutable
fields:~intuitively, only the last copy, called the \emph{master copy}, holds up
to date information.

The function~\cd{findMaster(obj)} returns a pointer to the master copy
of~\cd{obj}.
Intuitively, it simply has to walk the chain of forwarding pointers, starting
from~\cd{obj}.
However, while doing so it might encounter a forwarding pointer installed by a
promotion that is still ongoing.
In that case, we should wait for the promotion to complete.
We do so by acquiring the lock of the heap to which the copy belongs.
Since we acquire the lock in shared mode, we do not block concurrent calls to~\cd{findMaster}.
In contrast, promotion always locks in exclusive mode the heaps where it installs new forwarding pointers, ensuring mutual exclusion.

Our implementation of~\cd{findMaster} uses the classic double-checked locking pattern to reduce the cost of locking.
As long as we observe forwarding pointers, we move up, without locking~(l.~7).
Once we see an object that is a candidate for being the master copy, we acquire the lock of its heap, and check whether the object has acquired a forwarding pointer in the meantime; if not, it is definitely the master copy, and can be returned~(l.~8-9).
Otherwise, we unlock the heap and start walking the chain again~(l.~10).
Note that it is the caller's responsibility to unlock the heap.

\paragraph{Reading Mutable Data.}

Mutators can read a mutable field in an object by calling~\cd{readMutable(obj,field)}.
It would be correct to simply acquire the master copy, read the field, and release the lock~(l.~14-17).
We add a fast path:~read the mutable field optimistically, then check for the absence of a forwarding pointer~(l.~12-13).
This way, accessing mutable fields in objects without copies only takes a couple of machine instructions.

\paragraph{Writing Non-pointer Data.}

Writing plain data such as integers or floating-pointer numbers cannot involve promotion, and thus is always relatively cheap.
The implementation of~\cd{writeNonptr} mimicks that of~\cd{readMutable}.
In the fast path, we optimistically write the value~\cd{val} in~\cd{field}, and then check whether~\cd{obj} was the master copy~(l.~19-20).
Otherwise, we find the master copy, write the value, and unlock~(l.~21-23).

\paragraph{Writing Pointer Data.}

Writing pointer data is the most difficult case, as it might trigger a promotion.
The code for~\cd{writePtr(obj,field,ptr)}, which attempts to write~\cd{ptr} to~\cd{field} in~\cd{obj}, is given in~\figref{alg-2}.
We may have to promote~\cd{ptr} to the heap of~\cd{obj} if writing it directly would result in entanglement.
The algorithm can be decomposed into three cases:~a fast path, non-promoting writes, and promoting writes.
Let us describe each of these in turn.

The fast path of~\cd{writePtr}~(l.~2-5) writes~\cd{ptr} into~\cd{obj} only if the latter has no forwarding pointer and is in the heap of the currently running task.
Since this heap is necessarily a leaf in the hierarchy, promotion is never needed in this case.
This qualifies as a fast path since testing whether an object pointer belongs to the currently running task can be implemented much more efficiently than computing the depth of an arbitrary heap.

On the slow path, we acquire the master copy of~\cd{obj} and obtain the depth of its heap.
If it is deeper in the hierarchy than that of~\cd{ptr}, we are not creating entanglement, and can simply perform the write and unlock~(l.~6-10).
Otherwise, we have to promote, and thus unlock and call the dedicated function~\cd{writePromote(obj,field,ptr)}~(l. 11-12).

\begin{figure*}
  \newcommand{\dcheckmark}{\ensuremath{\checkmark\hspace{-.5em}\checkmark}}
  \centering
  \begin{tabular}{lccccc}
    &
    \multicolumn{2}{c}{Read} &
    \multicolumn{3}{c}{Write}
    \\
    \cmidrule(lr){2-3}\cmidrule{4-6}
    & Immutable
    & Mutable
    & Non-pointer
    & Non-promoting
    & Promoting
    \\\midrule
    Local
    & $\checkmark$
    & $\dcheckmark$
    & $\dcheckmark$
    & $\dcheckmark$
    &
    \\
    Distant
    & $\checkmark$
    & $\dcheckmark$
    & $\dcheckmark$
    & $\sim$
    & $\approx$
    \\
    Promoted
    & $\checkmark$
    & $\sim$
    & $\sim$
    & $\sim$
    & $\approx$
  \end{tabular}
  \hspace{.5cm}
  \begin{tabular}{|cl|}
    \multicolumn{2}{c}{}
    \\
    \hline
    $\checkmark$
    & single instruction
    \\
    $\dcheckmark$
    & few instructions
    \\
    $\sim$
    & single-heap locking
    \\
    $\approx$
    & path locking + copying
    \\
    \hline
  \end{tabular}
  \caption{Costs of memory operations.}
  \label{fig:costs}
\end{figure*}

\paragraph{Promoting Writes.}

We proceed in three phases.
First, we lock in exclusive mode all the heaps on the path from~\cd{ptr} to the master copy of~\cd{obj} from the bottom up~(l.~15-23).
Then, we promote~\cd{obj} and write the address of the promoted copy to~\cd{field}~(l.~24-25).
Finally, we unlock the path from top to bottom~(l.~26-27).

Acquiring the locks serve several purposes.
Let us call~$h_1$ the heap that contains~\cd{obj} and~$h_2$ the heap containing the master copy of~\cd{obj}.
By acquiring the locks on the path from~$h_1$ to~$h_2$ excluded, we take ownership of the forwarding pointer of any object we might need to promote.
By acquiring the lock of~$h_2$, we ensure that no concurrent call to~\cd{findMaster} will return before we have finished promoting.

The recursive function~\cd{promote} returns a pointer to a copy of~\cd{obj} held in~\cd{heap} or one of its parents.
If~\cd{obj} already resides in~\cd{heap} or above, it can simply be returned~(l.~29); if~\cd{obj} has a forwarding pointer, we follow it~(l.~30-31).
Taken together, these two tests ensure that objects with a chain of forwarding pointers leading above~\cd{heap} are not copied.
If both fail, we have to introduce a new copy.
We allocate a copy~\cd{newObj} of~\cd{obj} in~\cd{heap}~(l. 38), set the forwarding pointer slot of~\cd{obj} to point to~\cd{newObj}~(l.~32-33), and copy non-pointer fields~(l.~34-36).
We recursively promote pointer fields, since they might point strictly below~\cd{heap}~(l.~37-39).
At this point we can return~\cd{newObj}~(l.~40) since it belongs to~\cd{heap}, as do every object reachable from it.

Note that while we have expressed~\cd{promote} as a recursive function for simplicity, it can be implemented using a work list.
In particular, we were careful to set the forwarding pointer of~\cd{obj} before the recursive calls.

\paragraph{Cost Summary.}

\figref{costs} summarizes the costs of each memory operation in various situations.
Columns correspond to distinct memory operations, distinguishing between non-promoting and promoting pointer writes.
Rows classify objects being read from or written to:~\emph{promoted} objects are those with forwarding pointers;~\emph{local} and~\emph{distant} objects have no forwarding pointers and belong either to the heap of the task performing the memory operation~(local objects) or to one of its ancestors~(distant objects).


\subsection{Promotion-Aware Copy Collection}
\label{sec:promo-aware-collection}

Promotion introduces redundant copies of objects.
However, it is not difficult to eliminate these copies by piggybacking on the classic semispace garbage collection algorithm.
\iffull
We briefly explain our proposal; precise details are available in~Appendix~\ref{app:alg}.
\else
We briefly explain our proposal; precise details are available in the
technical appendix~\cite{this-paper-online}.
\fi

Assume that we are collecting a sub-tree starting at some heap~$h$; each heap
below~$h$~(included) thus acquires a to-space.
When examining an object in from-space, our collector traverses its forwarding
pointer chain, considering several possibilities in turn.
\begin{enumerate}
\item
  If the chain leads into a to-space, it points to a copy introduced during
  collection.
\item
  If the chain leads into a from-space that is strictly above~$h$ in the
  hierarchy, it leads to a copy introduced during promotion.
\item
  If the chain ends at an object that has no forwarding pointer, then
  this object is in a from-space below~$h$.
\end{enumerate}
In the first two cases, the address of the copy can be reused directly, whereas
in the last case we have to introduce a new copy.
The second case corresponds to the elimination of duplicates introduced during
promotion.
Note that since we do not attempt to access copies outside of the collection
zone, no additional locking is needed.
In the third case, we copy the last element of the forwarding chain
into the to-space and update its forwarding pointer to point to this
new copy.
By doing so, we effectively make sure that all pointers to the object
or its promoted copies will point to the new copy created in the
to-space.


\section{Implementation and Experiments}
\label{sec:exp}

We implemented our techniques by building upon the parallel MLton compiler
developed in prior work~\cite{hiheap-2016}, from which we inherit the
hierarchical heaps infrastructure, garbage collection policy, and scheduler.
We extended this compiler with support for general mutation by closely
following the algorithms described in~\secref{alg}.
\iffull
Further details on the implementation can be found in~Appendix~\ref{app:imp}.
\else
Further details on the implementation can be found in the
technical appendix~\cite{this-paper-online}.
\fi

We evaluate our techniques by considering a number of benchmarks
compiled with several compilers for dialects of parallel ML.
Thanks to the shared ML language, our benchmarks remain mostly
identical across compilers except for minor compatibility edits.
Our benchmark suite builds on previous suites from parallel functional
languages~\cite{manticore-implicit-11,abfr-manticore-gc-11}, and
extends them, also to include imperative programs which use mutable
data.
Our benchmarks use several standard implementations of data types such as sequences 
and graphs.
Unless stated otherwise, the elements of the sequences are 64-bit numeric types
(integers or floating point) generated randomly with a hash function.

\paragraph{Experimental Setup.}
For the measurements, we use a 72-core (4 x 18 core Intel E7-8867 v4)
Dell PowerEdge Server with 1 Terabyte of memory.
For the sequential baselines, we use the whole-program optimizing
MLton compiler~\cite{mlton}, which we label \cd{mlton}.
We compare all of our benchmarks to the work of Blelloch, Spoonhower,
and Harper~\cite{sbhg-space-10}, labeled \cd{mlton-spoonhower}, which
extends the MLton compiler to support nested fork-join parallelism and
parallel allocation, but utilizes sequential, stop-the-world collection.
For purely functional benchmarks, we also compare with the Manticore
compiler~\cite{manticore-implicit-11,abfr-manticore-gc-11}, labeled
\cd{manticore}, which provides for parallel functional programming by
using syntax similar to ours.
We refer to our hierarchical-heaps compiler as \cd{mlton-parmem}.

When taking timing measurements, we exclude initialization times.
All reported timings include GC times and are reported as the median
of five runs.

\newcommand{\bench}[1]{\cd{#1}}

\subsection{Pure Benchmarks}
These benchmarks are purely functional, meaning that their source code
does not use mutation.

\paragraph{\bench{fib}.}
  This benchmark computes the $42^\text{nd}$ Fibonacci number via
  the na\"ive recursive formula $F(n) = F(n-1) + F(n-2)$, with a sequential
  threshold of $n = 25$.

\paragraph{\bench{tabulate}, \bench{map}, \bench{reduce}, \bench{filter}.}
  These benchmarks each begin by generating an input sequence
  of size $10^8$.
  The \bench{tabulate} benchmark completes once the input sequence is built.
  The \bench{map} benchmark constructs a second sequence by applying a simple
  function to each element.
  The \bench{reduce} benchmark sums the elements of the input sequence.
  The \bench{filter} benchmark constructs a second sequence containing only
  the elements which satisfy a given predicate.
  They are each implemented with straightforward divide-and-conquer approaches,
  with a sequential threshold of $10^4$ elements.

\paragraph{\bench{msort-pure}.}
  This benchmark first tabulates a 
  sequence of size
  $10^7$. It then sorts the sequence with a function
  similar to the one shown in \figref{msort-imperative}, except that it uses a
  purely functional quick-sort below a sequential threshold of $10^4$
  instead of the imperative one.

\paragraph{\bench{dmm}, \bench{smvm}.}
  These benchmarks operate on square matrices of size $n\times n$.
  Each matrix is represented by
  a sequence of rows (or columns).
  The \bench{dmm} benchmark multiplies two dense matrices with the na\"ive $O(n^3)$ algorithm,
  where each of the $n$ rows is implemented as another 
  sequence of size $n$.
  The \bench{smvm} benchmark multiplies a sparse matrix by a dense vector, where each row
  of the sparse matrix contains only the non-zero entries represented as
  index-value pairs.
  In \bench{dmm}, $n = 600$. In \bench{smvm}, $n = 20,000$ and each row
  has approximately $2,000$ non-zero entries.
  The sequential threshold is one matrix row.

\paragraph{\bench{strassen}.}
  This benchmark multiplies two dense square matrices of size $n \times n$ using
  Strassen's algorithm.
  The matrices are represented by quadtrees with leaves of
  vectors of elements of size $64\times64$.
  In our experiments, $n = 1024$ with a sequential threshold of 64 (that is,
  the leaves of the quadree are each processed sequentially).

\paragraph{\bench{raytracer}.}
  This benchmark is adapted from the raytracer benchmark
  written for the Manticore language~\cite{abfr-manticore-gc-11},
  which was adapted from an Id program~\cite{nikhil91a}.
  It renders a $600\textrm{px} \times 600\textrm{px}$ scene in parallel by
  tabulating a 
  sequence of pixels with a sequential granularity of 300 pixels.

\subsection{Imperative benchmarks.}
These benchmarks are designed to exercise various different forms of
mutation, as summarized in~\figref{bench-representative-op}.
Due to mutation, they are not implementable in Manticore.

\paragraph{\bench{msort}, \bench{dedup}.}
  These benchmarks begin by tabulating a 
  sequence of
  size $10^7$ before sorting it with a technique similar to that shown in
  \figref{msort-imperative}.
  The \cd{msort} benchmark uses imperative quick-sort below the sequential
  threshold of $10^4$ elements.
 The \bench{dedup} benchmark is similar to \bench{msort} but  removes duplicate keys.
  Below the sequential threshold, this is accomplished by imperatively
  inserting elements into a hash set before sorting with the in-place quick-sort.
  For \bench{dedup}, we guarantee the sequence has approximately $10^6$ unique keys.

\paragraph{\bench{tourney}.}
  This benchmark tabulates a 
  sequence of
  $10^8$ contestants, and then computes a tournament tree.
  Each contestant is represented by an integer which measures their fitness.
  In the tournament tree, each contestant $c$ has an associated parent pointer
  which points to the contestant that eliminated $c$ from the tournament. 
  This benchmark computes the tournament tree with a simple divide-and-conquer
  approach, using mutation at each join point in order to set a parent pointer.

\paragraph{\bench{usp}, \bench{usp-tree}, \bench{multi-usp-tree}, \bench{reachability}.}
These benchmarks consider variants of parallel breadth-first
search (BFS) on directed, unweighted graphs.
BFS visits vertices in rounds.
At round $r$, BFS visits (in parallel) every vertex which has not
previously been visited and is reachable by $r$ hops from the source
vertex.
When a vertex is visited, a piece of mutable data associated with it
is updated.

The BFS variants differ in the types of per-vertex mutable data.
They also differ in the number of times a vertex is visited.
Except in the \bench{reachability} benchmark, we guarantee that each
vertex is visited exactly once by marking vertices as visited with an
atomic ``compare-and-swap'' operation.
In the \bench{reachability} benchmark, we check and update the visited
status of vertices simply by reading and writing to a shared flag.
This creates a data race and potentially causes some vertices to be visited
multiple times (up to at most $P$ visitations, $P$ = number of processors).
In practice, this variant of BFS often performs better on modern hardware because
(a) atomic operations such as compare-and-swap are expensive, and (b) observing
the data race within a particular execution is rare.

The specifics of these benchmarks are described below.
\begin{itemize}
\item \textbf{\bench{reachability}} identifies which vertices are reachable
  from the source.
\item \textbf{\bench{usp}} computes the unweighted single-source shortest path
  length of all vertices.
  Every time a vertex is visited, the algorithm records the current round number
  as the distance to that vertex.
\item \textbf{\bench{usp-tree}} computes all unweighted single-source
  shortest paths.
  It is implemented with an array $A$ of ancestor lists.
  When a vertex $v$
  is visited along an edge $(u,v)$, the ancestors of $v$ are recorded as
  \mbox{$A[v] := u :: A[u]$}.
\item {\textbf{\bench{multi-usp-tree}}} runs 36 copies of~\bench{usp-tree} in
  parallel.
\end{itemize}
The input graph is the \emph{orkut} social network
graph~\cite{stanford-network-data}, which has approximately 3 million
vertices, 117 million edges, and a diameter of 9.
Each benchmark begins by converting the input graph into a
compact adjacency-sequence format suitable for parallel BFS.

\subsection{Representative Operations}

In~\figref{bench-representative-op}, using the terminology from~\figref{costs},
we characterize each benchmark by a \emph{representative memory operation}.
Representative operations summarize which type
of operation is most likely to be a dominant cost in execution time.%
\footnote{Note that immutable reads are pervasive in all benchmarks.}
These in turn help understand and predict performance.
For example, if a benchmark exhibits mostly ``local, non-pointer'' writes, then
that benchmark likely has low overhead; in contrast, if a benchmark is
characterized by many ``distant, promoting'' writes, then it might have high
overhead and scale poorly.

\begin{figure}
\begin{tabular}{rl}
\toprule
Benchmark & Representative Operation \\\midrule
pure benchmarks & immutable reads \\
\cd{msort} & local non-pointer writes \\
\cd{dedup} & local non-pointer writes \\
\cd{tourney} & local non-promoting writes \\
\cd{reachability} & distant non-pointer writes \\
\cd{usp} & distant non-pointer writes \\
\cd{usp-tree} & distant promoting writes \\
\cd{multi-usp-tree} & distant promoting writes \\\bottomrule
\end{tabular}
\caption{Representative operations of all benchmarks.}
\label{fig:bench-representative-op}
\end{figure}

\subsection{Results}
\label{sec:results}
\begin{figure*}
{
\small
\centering

\begin{tabular}{r*{16}c}
%
%
&
\multicolumn{2}{c}{\cd{mlton}} &
\multicolumn{5}{c}{\cd{mlton-spoonhower}} &
\multicolumn{4}{c}{\cd{manticore}} &
\multicolumn{5}{c}{\cd{mlton-parmem} (our compiler)} \\
\cmidrule(lr){2-3}\cmidrule(lr){4-8}\cmidrule(lr){9-12}\cmidrule(lr){13-17}
&
$T_s$ & $GC_s$ &
$T_1$ & {\scriptsize\begin{tabular}[c]{@{}c@{}}Over\\head\end{tabular}} & $T_{72}$ & {\scriptsize\begin{tabular}[c]{@{}c@{}}Speed\\up\end{tabular}} & $GC_{72}$ &
$T_1$ & {\scriptsize\begin{tabular}[c]{@{}c@{}}Over\\head\end{tabular}} & $T_{72}$ & {\scriptsize\begin{tabular}[c]{@{}c@{}}Speed\\up\end{tabular}} &
$T_1$ & {\scriptsize\begin{tabular}[c]{@{}c@{}}Over\\head\end{tabular}} & $T_{72}$ & {\scriptsize\begin{tabular}[c]{@{}c@{}}Speed\\up\end{tabular}} & $GC_{72}$
\\
\midrule
\cd{fib}             & 2.67   & 0.0\%  & 3.72   & 1.39   & 0.12   & 22.25  & 3.4\%  & 5.19   & 1.94   & 0.12   & 22.25  & 3.63   & 1.36   & 0.07   & 38.14  & 0.0\%  \\
\cd{tabulate}        & 1.11   & 39.4\% & 0.89   & 0.8    & 0.42   & 2.64   & 89.5\% & 1.92   & 1.73   & 0.12   & 9.25   & 1.62   & 1.46   & 0.07   & 15.86  & 15.4\% \\
\cd{map}             & 1.46   & 29.8\% & 1.31   & 0.9    & 0.48   & 3.04   & 82.2\% & 4.02   & 2.75   & 0.49   & 2.98   & 2.75   & 1.88   & 0.14   & 10.43  & 16.2\% \\
\cd{reduce}          & 1.13   & 40.5\% & 0.9    & 0.8    & 0.46   & 2.46   & 84.7\% & 1.93   & 1.71   & 0.13   & 8.69   & 1.43   & 1.27   & 0.09   & 12.56  & 11.7\% \\
\cd{filter}          & 3.62   & 11.9\% & 4.84   & 1.34   & 0.54   & 6.7    & 76.6\% & 6.15   & 1.7    & 0.49   & 7.39   & 5.75   & 1.59   & 0.18   & 20.11  & 12.0\% \\
\cd{msort-pure}      & 7.02   & 12.3\% & 8.52   & 1.21   & 1.79   & 3.92   & 75.8\% & --     & --     & --     & --     & 6.91   & 0.98   & 0.36   & 19.5   & 15.7\% \\
\cd{dmm}             & 3.76   & 15.2\% & 7.02   & 1.87   & 0.92   & 4.09   & 78.3\% & 8.3    & 2.21   & 0.19   & 19.79  & 5.83   & 1.55   & 0.16   & 23.5   & 6.8\%  \\
\cd{smvm}            & 7.23   & 0.0\%  & 9.93   & 1.37   & 0.2    & 36.15  & 0.0\%  & 12.68  & 1.75   & 0.32   & 22.59  & 8.69   & 1.2    & 0.18   & 40.17  & 0.0\%  \\
\cd{strassen}        & 2.54   & 1.6\%  & 2.89   & 1.14   & 0.16   & 15.88  & 40.6\% & 4.36   & 1.72   & 0.12   & 21.17  & 2.94   & 1.16   & 0.12   & 21.17  & 7.9\%  \\
\cd{raytracer}       & 7.41   & 1.3\%  & 7.0    & 0.94   & 0.3    & 24.7   & 29.8\% & 6.97   & 0.94   & 0.17   & 43.59  & 6.52   & 0.88   & 0.12   & 61.75  & 0.5\%  \\
\end{tabular}
}
\caption{%
Execution times (in seconds), overheads, and speedups of purely functional benchmarks.
}
\label{fig:pure-bench}
\end{figure*}

\begin{figure*}
\small
\centering

\begin{tabular}{r*{12}{c}}
%
%
&
\multicolumn{2}{c}{\cd{mlton}} &
\multicolumn{5}{c}{\cd{mlton-spoonhower}} &
\multicolumn{5}{c}{\cd{mlton-parmem} (our compiler)} \\
\cmidrule(lr){2-3}\cmidrule{4-8}\cmidrule(lr){9-13}
&
$T_s$ & $GC_s$ &
$T_1$ & {\scriptsize\begin{tabular}[c]{@{}c@{}}Over\\head\end{tabular}} & $T_{72}$ & {\scriptsize\begin{tabular}[c]{@{}c@{}}Speed\\up\end{tabular}} & $GC_{72}$ &
$T_1$ & {\scriptsize\begin{tabular}[c]{@{}c@{}}Over\\head\end{tabular}} & $T_{72}$ & {\scriptsize\begin{tabular}[c]{@{}c@{}}Speed\\up\end{tabular}} & $GC_{72}$
\\
\midrule
\cd{msort}           & 3.75   & 3.4\%  & 4.66   & 1.24   & 0.36   & 10.42  & 58.3\% & 5.33   & 1.42   & 0.18   & 20.83  & 7.7\%  \\
\cd{dedup}           & 3.72   & 2.6\%  & 4.05   & 1.09   & 0.32   & 11.63  & 52.1\% & 4.61   & 1.24   & 0.16   & 23.25  & 6.3\%  \\
\cd{tourney}         & 4.64   & 6.3\%  & 8.17   & 1.76   & 0.92   & 5.04   & 76.2\% & 7.86   & 1.69   & 0.23   & 20.17  & 7.1\%  \\
\cd{reachability}    & 8.36   & 0.0\%  & 21.59  & 2.58   & 0.52   & 16.08  & 0.0\%  & 19.59  & 2.34   & 0.46   & 18.17  & 0.0\%  \\
\cd{usp}             & 8.34   & 0.0\%  & 23.38  & 2.8    & 0.61   & 13.67  & 0.0\%  & 21.85  & 2.62   & 0.58   & 14.38  & 0.0\%  \\
\cd{usp-tree}        & 8.63   & 0.0\%  & 23.79  & 2.76   & 0.63   & 13.7   & 0.0\%  & 22.3   & 2.58   & 7.93   & 1.09   & 0.0\%  \\
\cd{multi-usp-tree}  & 100.25 & 3.0\%  & 209.59 & 2.09   & 19.07  & 5.26   & 34.0\% & 245.6  & 2.45   & 11.18  & 8.97   & 8.2\%  \\
\end{tabular}
\caption{%
Execution times (in seconds), overheads, and speedups of imperative benchmarks.
}
\label{fig:imperative-bench}
\end{figure*}

\begin{figure}
\centering
\includegraphics[width=.98\columnwidth]{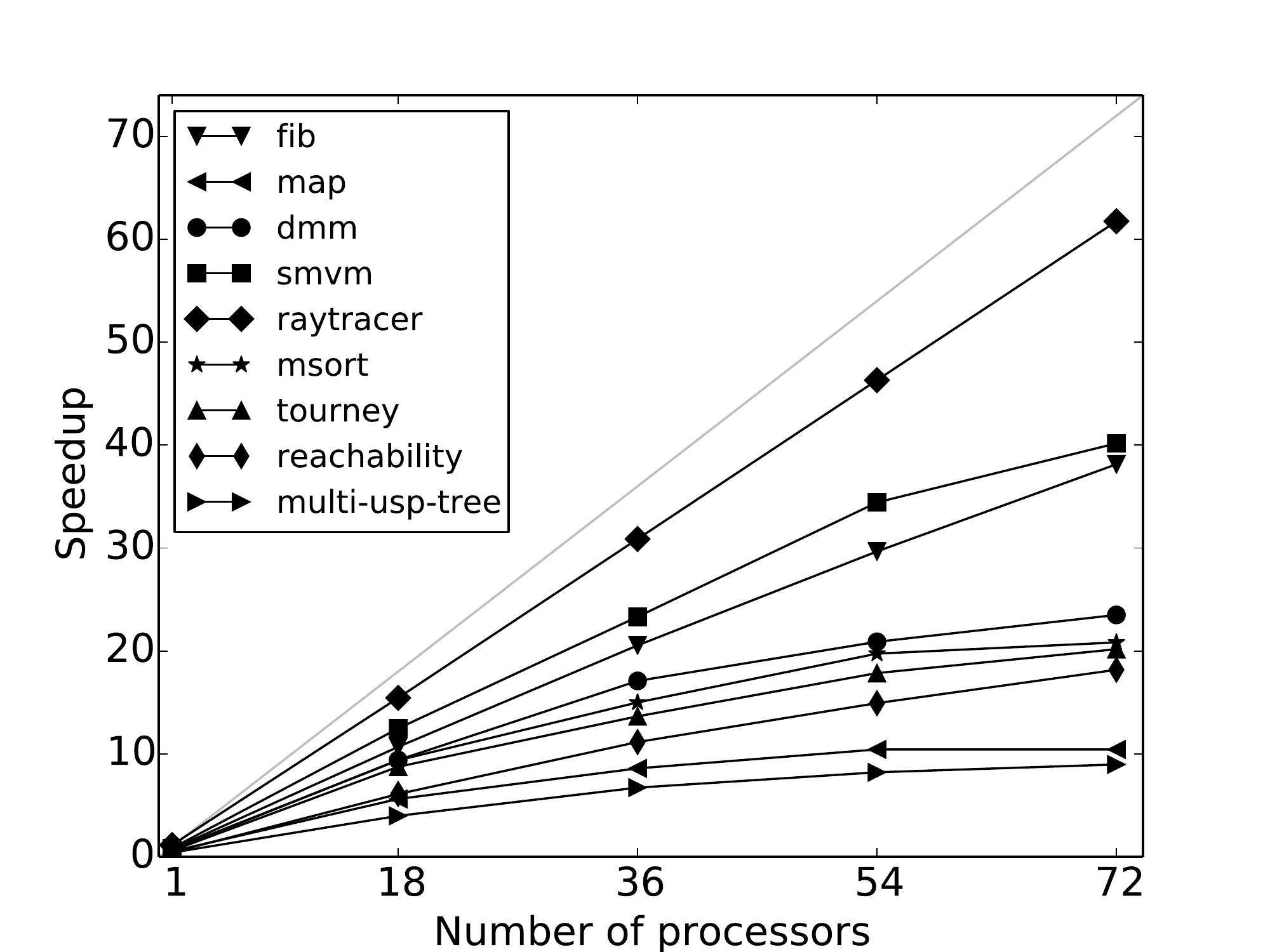}
\caption{Speedups of \cd{mlton-parmem}.}
\label{fig:speedup-plot}
\end{figure}

\begin{figure}
{
\small
\begin{tabular}{rccccc}
&
\cd{mlton} &
\multicolumn{2}{c}{\begin{tabular}[c]{@{}c@{}}\cd{mlton-}\\\cd{spoonhower}\end{tabular}} &
\multicolumn{2}{c}{\begin{tabular}[c]{@{}c@{}}\cd{mlton-parmem}\\(our compiler)\end{tabular}} \\
\cmidrule(lr){2-2}\cmidrule(lr){3-4}\cmidrule(lr){5-6}
        & $M_s$ & $I_1$ & $I_{72}$ & $I_1$ & $I_{72}$ \\
\midrule
\cd{fib}             & 0.0    & +0.0   & +0.31  & +0.0   & +0.4   \\
\cd{tabulate}        & 3.48   & 0.23   & 0.34   & 0.97   & 1.25   \\
\cd{map}             & 1.6    & 1.01   & 1.42   & 3.07   & 5.07   \\
\cd{reduce}          & 0.8    & 1.01   & 2.02   & 3.15   & 4.95   \\
\cd{filter}          & 2.8    & 0.8    & 1.41   & 1.7    & 2.51   \\
\cd{msort-pure}      & 1.43   & 1.01   & 1.44   & 1.49   & 4.9    \\
\cd{dmm}             & 0.18   & 1.78   & 3.39   & 0.94   & 7.39   \\
\cd{smvm}            & 1.92   & 1.33   & 1.47   & 2.67   & 2.78   \\
\cd{strassen}        & 0.22   & 0.95   & 4.41   & 1.86   & 17.68  \\
\cd{raytracer}       & 0.13   & 1.15   & 3.77   & 1.46   & 5.38   \\
\midrule
\cd{msort}           & 1.09   & 0.78   & 1.5    & 1.6    & 4.04   \\
\cd{dedup}           & 0.56   & 1.18   & 2.61   & 1.63   & 6.48   \\
\cd{tourney}         & 0.8    & 6.24   & 8.34   & 6.91   & 8.66   \\
\cd{reachability}    & 3.87   & 1.01   & 2.29   & 1.5    & 2.15   \\
\cd{usp}             & 3.87   & 1.01   & 2.35   & 1.5    & 2.19   \\
\cd{usp-tree}        & 3.97   & 1.01   & 2.25   & 1.55   & 2.21   \\
\cd{multi-usp-tree}  & 19.76  & 1.2    & 2.85   & 1.64   & 1.79   \\
\end{tabular}
}
\caption{Memory consumption (in GB) and inflations.}
\label{fig:bench-space}
\end{figure}

We collect the following statistics for each benchmark.
\begin{itemize}
  \item $T_s$ is the sequential execution time.
  \item $T_1$ and $T_{72}$ are execution times on 1 and 72 processors.
  \item The \emph{overhead} is $T_1 / T_s$.
  \item The \emph{speedup} is $T_s / T_{72}$. In general, the speedup on $P$
    processors is given by $T_s / T_P$.
  \item $GC_s$ is the percent of time spent in GC during a sequential run.
  \item $GC_{72}$ is the percent of processor time spent in GC during a
    $72$-processor run.
  \item $M_s$ is the memory consumption of the sequential run.
  \item $I_1$ and $I_{72}$ are memory inflations on 1 and 72 processors.
\end{itemize}
The \emph{memory consumption} statistic is an upper bound on the amount of physical memory
required to store heap-allocated objects; it is computed by tracking the
maximum heap occupancy within one execution, and includes fragmentation due
to parallel allocations.
\emph{Memory inflation} gives the memory consumption as a factor relative to
the sequential memory consumption, $M_s$. 

For \cd{mlton-spoonhower}, $GC_{72}$ includes processor time spent blocked
during a stop-the-world collection.
We do not report GC statistics for Manticore, because it is not able
to collect statistics only for a specific region of code, which we need to
have a meaningful comparison.
We are also unable to report results for Manticore on \bench{msort-pure}
due to a compiler bug.

Our results are summarized in Figures \ref{fig:pure-bench},
\ref{fig:imperative-bench}, \ref{fig:speedup-plot}, and \ref{fig:bench-space}.
Figures \ref{fig:pure-bench} and \ref{fig:imperative-bench} list the execution
times, overheads, speedups, and GC percentages of pure and imperative benchmarks,
respectively.
Figure \ref{fig:speedup-plot} shows the speedup of \cd{mlton-parmem} on 
various benchmarks for processor counts between 1 and 72.
Finally, Figure \ref{fig:bench-space} lists the memory consumptions and
inflations of all benchmarks.

\paragraph{Overheads.}
Inspecting \figreftwo{pure-bench}{imperative-bench}, we can
make several observations and conclusions.
First, for both pure and imperative benchmarks, the overheads of
\cd{mlton-parmem} are generally comparable to those of \cd{mlton-spoonhower},
which serves as a good baseline because it does not support parallel
memory management.
This shows that our techniques for maintaining a dynamic memory
structure based on hierarchical heaps can be implemented
efficiently.
Second, we observe that for pure benchmarks (\figref{pure-bench}), our
overheads are within a factor 2 of the sequential baseline and are
consistently smaller than those of \cd{manticore}.
Third, for imperative benchmarks (\figref{imperative-bench}), our overheads are
higher than those of the pure benchmarks but still remain within a factor
of approximately 2.6 in comparison to the sequential baseline.
The increase in overheads is due in part to the memory operations
which are no longer plain loads and stores~(see~\figreftwo{costs}{bench-representative-op}).

\paragraph{Speedups.}
Inspecting \figref{pure-bench}, we
observe that for pure benchmarks with 72 cores, \cd{mlton-parmem} achieves
speedups ranging between 10 and 62.
Compared to \cd{mlton-spoonhower}, our speedups are significantly
higher, which is expected because \cd{mlton-spoonhower} suffers from
sequential, stop-the-world garbage collections.
Our speedups are also significantly better than those of
\cd{manticore}, which is sometimes as low as 3x.
This seems surprising because \cd{manticore} is designed for purely
functional programs.
The reason is that \cd{manticore} relies on imperative updates within the
run-time to execute computations in parallel (e.g., to communicate the result
of a remotely executed task to another processor) and employs a promotion
technique to preserve local heap invariants~\cite{abfr-manticore-gc-11}.
To verify this, we measured that on the \bench{map} benchmark with 72
cores, \cd{manticore} promoted nearly 340MB of data in total, whereas
\cd{mlton-parmem} performed no promotions.

Inspecting the imperative benchmarks (\figref{imperative-bench}), we
observe that \cd{mlton-parmem} achieves good speedups on 72 processors
(between 14 for high-overhead benchmarks and 23 for those with
low overhead) with a couple exceptions: the highly concurrent
\bench{usp-tree} and \bench{multi-usp-tree} benchmarks.
Poor performance on \bench{usp-tree} and \bench{multi-usp-tree} is
expected, because these benchmarks exhibit close to pessimal cases for
our techniques with frequent concurrent updates of shared pointer data.
For example, every time a vertex is visited in \bench{usp-tree}, one cell
of a distant array (located at the root) is updated with a new list, triggering
promotion from a leaf heap to the root heap.
Since promotions require locking entire heaps, they can sequentialize
otherwise parallel visitations, leading effectively to complete
serialization of the entire computation.
However, when multiple \bench{usp-tree} computations are performed in parallel in
the \bench{multi-usp-tree} benchmark, some promotions remain
independent and execute in parallel because the updated array is not
always at the root heap.
We indeed see a 9-fold speedup in this benchmark.

In \figref{speedup-plot}, we observe that as the number of processors
increases, the speedup of all benchmarks continues to increase.
That is, there are no inversions.
For multiple benchmarks the speedup improves nearly linearly, suggesting
the possibility of further scalability to higher core counts.

\paragraph{Garbage Collection.}
Inspecting \figreftwo{pure-bench}{imperative-bench}, we observe
that~\cd{mlton-parmem} only loses at most approximately
16\% of its time to garbage collection on runs with 72 cores.
As expected, \cd{mlton-spoonhower} performs poorly due to its
sequential GC.
Some benchmarks (\bench{smvm}, \bench{reachability}, \bench{usp},
\bench{usp-tree}) spend no time in garbage collection, regardless of when
run sequentially or in parallel.
This is due to the fact that these benchmarks allocate memory in an already large
heap, which was grown to accommodate the input.
(In the case of \bench{smvm}, the benchmark does not include input generation;
for the graph algorithms, the benchmark does not include the time it takes
to read the graph from disk into a large string in the heap.)

\paragraph{Memory Consumption.}
Inspecting~\figref{bench-space}, with only a few exceptions, our compiler on
72 cores consumes at most 7x more memory than the sequential baseline.
Note that in general, any $P$-processor execution scheduled via work-stealing
can expect to see inflation of up to a factor $P$~\cite{BlumofeLe98,BlumofeLe99}.
Thus, using \cd{mlton-spoonhower} as an alternative baseline helps determine
how much inflation is due simply to parallel execution, versus how much is due
to our techniques.
Indeed, our inflation with respect to \cd{mlton-spoonhower} is generally
lower, staying consistently within a factor of approximately 4.
Our implementation introduces additional inflation in part through (a) the need
for a separate forwarding pointer on every object, and (b) greater
fragmentation of allocation to distinguish heaps within the hierarchy.

\section{Discussion and Future Work}\label{sec:discussion}

The promotion techniques presented in this paper rely on coarse-grained
locks to manage concurrent manipulations of overlapping data.
This approach prevents certain promotions from proceeding in parallel, even
when those promotions would otherwise be independent.
For example, in the \cd{usp-tree} benchmark, every visitation of a vertex
triggers a promotion to the root of hierarchy, causing a serialization of
visitations.
However none of these promotions overlap, so they ought to be able to
proceed in parallel.
In future work, we intend to design a more fine-grained promotion
strategy that would permit parallel promotions to the same heap.

With respect to garbage collection, our current implementation has two
limitations: first, it can only collect leaf heaps; second,
each such collection is sequential.
Parallelism is thus achieved by collecting many leaves independently.
Our results suggest that this simple approach can perform well for highly
parallel applications.
However, if a collection takes place at the root heap (when there is no
parallelism), such a collection would be sequential and effectively
stop-the-world.
More generally, when there is little parallelism, large collections can
take place sequentially.
In future work, we plan to complete the implementation by adding support for
parallel collection of individual heaps, and in general parallel collection
of sub-trees of heaps (not just leaves).

\section{Related Work}
\label{sec:related}

With the proliferation of shared memory parallel computers using
modern multicore processors, there has been significant work on the
design and implementation of high-level programming languages for
writing parallel
programs~\cite{nesl-94,Lea00,TPL09,X10-2005,Keller+2010,manticore-implicit-11,szj-multimlton14,is-habanero-14,golang-2017}.
For implicit memory allocation and reclamation with garbage
collection, there are numerous techniques for incorporating
parallelism, concurrency, and real-time features. 
\citet{jone11} provides an excellent survey.

We contrast our work with a number of
systems~\cite{doli93,doli94,doma02,ande10,abfr-manticore-gc-11,marl11,szj-multimlton14,hiheap-2016}
that use processor- or thread-local heaps which service (most)
allocations of the executing computation and can be collected
independently combined with a shared global heap that must be
collected cooperatively.

The Doligez-Leroy-Gonthier (DLG) parallel
collector~\cite{doli93,doli94} employs this design, with the invariant
that there are no pointers from the shared global heap into any
processor-local heap and no pointers from one processor local-heap
into another processor-local heap.
To maintain this invariant, all mutable objects are allocated in the
shared global heap and (transitively reachable) data is promoted
(copied) from a processor-local heap to the shared global heap when
updating a mutable object, which increases the cost of all mutable
allocations and updates.
In our approach, mutable objects are allocated in the thread-local
heap and updates to such never need to promote data, making this
common case significantly less expensive than in DLG.
Moreover, in DLG, scheduling and communication actions, such as
migrating a language-level thread from one processor to another or
returning the result from a child task to a parent task, typically
employ mutable objects and require promotions.
With hierarchical heaps, the local heap is associated with the task,
rather than the processor, and returning the result of a child task to
the parent task is accomplished without copying.

Anderson~\cite{ande10} describes TCG, a variant of the DLG collector
for a language with implicit parallelism serviced by a fixed number of
worker threads pinned to processors.  TCG allows mutable objects to be
allocated in the processor-local heap and a processor-local collection
copies live data to the global shared heap.  When updating a mutable
object in the shared global heap with a processor-local pointer, a
processor-local garbage collection is triggered, which copies the
to-be-written object (and all other processor-local live data) to the
shared global heap; updating a mutable object in the processor-local
heap proceeds without a collection.  Using collection to
over-approximate promotion would not work well with our hiearchical
heaps, because the collection would need to be triggered for
\emph{all} descendent heaps of the mutable object being written and
could not be performed by the writing processor independently.

The Manticore garbage collector~\cite{abfr-manticore-gc-11} is another
variant of the DLG design, where the Appel semi-generational
collector~\cite{appe89} is used for collection of the processor-local
heaps.
Although the high-level language is mutation-free, the implementation
uses mutation to realize various parallel constructs and employs
promotion to preserve the heap invariants.
Recent work~\cite{le-fluet-icfp15} has considered extending the
Manticore language with mutable state via software transactional
memory, but notes that promotions make a chronologically-ordered read
set implemented as a mutable doubly-linked list inefficient.
In contrast, such a data structure would be efficient in our
hierarchical heaps, since a transaction's read set is necessarily
local to the thread executing the transaction.

The current Glasgow Haskell Compiler garbage collector~\cite{marl11}
combines elements of the DLG and Domani et al.~\cite{doma02}
collectors.  Although Haskell is a pure language, there is significant
mutation due to lazy evaluation.  The collector allows mutable objects
to be allocated in a dedicated portion of the processor-local heaps,
which use a non-moving collector.  The collector also allows pointers
from the global heap to the processor-local heaps, mediated by proxy
objects.  When another processor accesses a proxy, it communicates
with the owning processor to request that the object be promoted to
the global shared heap.  Proxy objects fit well with Haskell's lazy
evalution, where all pointer accesses have a read barrier to check for
unevaluated computations and where proxy objects can be incorporated
into unevaluated computations without requiring promotion.
Standard~ML employs strict evaluation and eager promotion is a better
fit.

The MultiMLton project~\cite{szj-multimlton14} forked from
\cd{mlton-spoonhower} and shifted the domain to message-passing
concurrency. While one could encode shared-state fork-join parallelism
with their message-passing operations, the resulting overheads
(emulating references with threads; eager thread creation vs lazy
work-stealing) would not lead to a meaningful performance comparison.
Their garbage collection strategy is tuned to their message-passing
concurrency bias --- starting from the DLG design and invariants, they
avoid promotion through procrastination, blocking the writing thread
(and executing another of the abundant concurrent threads) until a GC
can be performed, which promotes the object and fixes references.
They are concerned that promotion, leaving forwarding pointers that
overwrite object data, would have unacceptable read-barrier overhead,
which also motivates their dynamic cleanliness analysis; in contrast,
we have introduced a dedicated forwarding-pointer metadata field,
which only requires a read barrier for mutable data.

Raghunathan et al. introduced hierachical heaps~\cite{hiheap-2016} to
mirror the hierarchy of tasks in a strict pure functional language
with nested parallelism.  They prove that the language enforces
disentanglement, formulate a hierarchical garbage collection technique
that allows independent heaps to be collected concurrently, and report
the performance of an implementation in MLton.  We extend this work to
accomodate the mutable references and arrays of a strict impure
functional language with nested parallelism.
The hierarchical-heaps design is partly motivated by a desire to take
advantage of the natural data locality of
computation~\cite{abfmr-couple-15}, which, with some care, could be
preserved by thread schedulers (e.g.,~\cite{abb02,bfgs11}).

\section{Conclusion}

The high-level nature of functional programming languages makes them a
good fit for parallel programming, but they require sophisticated
memory managers which are challenging to get right in the joint
presence of parallelism and mutation.
In this paper, we showed how to provide efficient support for uses of
mutation common in parallel programs by exploiting the hierarchical
structure of functional computations.
Our experiments suggest that these results could be an important step
towards making functional programming a serious contender for
performant parallel computing.

\begin{acks}
  We thank Rohan Yadav for his assistance in the implementation of
  the benchmarks used in this paper.
  This material is based upon work supported by the
  \grantsponsor{GS100000001}{National Science
    Foundation}{http://dx.doi.org/10.13039/100000001} under Grant
  No.~\grantnum{GS100000001}{1408940}, Grant
  No.~\grantnum{GS100000001}{1408981}, and Grant No.~\grantnum{GS100000001}{1629444}.  Any opinions, findings, and
  conclusions or recommendations expressed in this material are those
  of the author and do not necessarily reflect the views of the
  National Science Foundation.
  The first author was also partially supported by
  the~\grantsponsor{GS501100001659}{German Research
    Council~(DFG)}{http://dx.doi.org/10.13039/501100001659} under Grant
  No.~\grantnum{GS501100001659}{ME14271/6-2}.
\end{acks}

\newpage
{
\iffull
\bibliography{main,new,gc-2016,thisbib}
\else
\bibliography{../../bibliography/main,../../bibliography/new,../../bibliography/gc-2016,thisbib}
\fi
}

\iffull
\clearpage
\appendix
\section{Algorithms}
\label{app:alg}

\subsection{Promotion-Aware Copy Collection}

\begin{figure}
\begin{lstlisting}[language=pseudocode]
function toSpaceOf: heap -> heap
function isToSpace: heap -> bool
function switchSemispaces: heap -> unit
\end{lstlisting}
\hspace{.5cm}
\begin{lstlisting}[language=pseudocode]
function collect (topHeap) =
  for r in current roots:
    *r $\la$ cheneyCopy(heap, *r)
  for h below topHeap included:
    switchSemispaces(h)
function cheneyCopy (topHeap, obj) =
  heap $\la$ heapOf(obj)
  if depth(heap) $<$ depth(topHeap): return obj
  if isToSpace(heap): return obj
  if hasFwdPtr(obj):
    return cheneyCopy(topHeap, *fwdPtr(obj))
  newObj $\la$ freshObj(toSpaceOf(heap), sizeOf(obj))
  *fwdPtr(obj) = newObj
  for field in nonptrFields(obj):
    *getField(newObj, field) $\la$
      *getField(obj, field)
  for field in ptrFields(obj):
    *getField(newObj, field) $\la$
      cheneyCopy(topHeap, *getField(obj, field))
  return newObj
\end{lstlisting}
\caption{Promotion-aware copy collection.}
\label{fig:alg-gc}
\end{figure}

Promotion introduces redundant copies of memory objects.
We now present a way to eliminate these copies by piggybacking on the classic semispace garbage collection algorithm.
Our proposal is given in~\figref{alg-gc}, including additional primitives specific to semispace collection.

We now assume that every hierarchical heap accessed by the mutator is paired with another heap used only during collection.
Following standard terminology, we call the former a~\emph{from-space} and the latter a~\emph{to-space}.
Our collection algorithm manipulates semispaces using three primitives:~\cd{toSpaceOf} returns the to-space associated with a given from-space;~\cd{isToSpace(heap)} returns true iff~\cd{heap} is a to-space;~\cd{switchSemispaces} swaps to-space and from-space.

The function~\cd{collect(topHeap)} is called to collect the subtree starting at~\cd{topHeap}.
We assume that every task associated with a leaf heap below~\cd{topHeap} has been suspended by the runtime system.
Thanks to disentanglement, this is sufficient to collect the entire subtree independently from other mutators and collectors.
It copies every object reachable from a root to the to-spaces using~\cd{cheneyCopy}~(l. 2-3) and then swaps the semispaces of every heap in the subtree~(l.~4-5).

The function~\cd{cheneyCopy} takes a from-heap~\cd{topHeap} and an object pointer~\cd{obj}, and returns a copy of~\cd{obj}.
This copy is guaranteed to be either in a to-heap below~\cd{topHeap}, or in a from-heap strictly above~\cd{topHeap}.
The latter case corresponds to the elimination of copies introduced during promotion:~copy collection replaces a pointer to an old copy with a pointer to the a more recent one lying outside of the collection zone.
In addition, since we do not follow forwarding pointers that belong to objects outside of the collection zone, we do not have to lock heaps during collection.

Like~\cd{promote}, we specify~\cd{cheneyCopy} as a recursive function.
Let us call~\cd{heap} the heap where~\cd{obj} resides.
If~\cd{heap} is strictly above~\cd{topHeap}~(l.~9-10), or is a to-space~(l.~11-12),~\cd{obj} can be returned.
If~\cd{obj} has a forwarding pointer,~\cd{cheneyCopy} follows it~(l.~13-14).
Otherwise, as in~\cd{promote}, we create a new copy, setting up the forwarding pointer of~\cd{obj} to point to it, recursively call~\cd{cheneyCopy} on its pointer fields, and return it~(l.~17-23).

\section{Implementation}
\label{app:imp}
\label{app:impl}

\paragraph{Scheduler.}
Any implementation of a fork/join programming model requires a
scheduler to coordinate work between worker threads. Tasks are
evaluated within a ``user-level thread'' that is scheduled onto a
``worker thread''. In addition to the \cd{forkjoin} function
exposed to the mutator, the scheduler also exposes a \cd{schedule}
function to idle worker threads to find waiting work.

A naive implementation of a work-stealing scheduler will create tasks
for both thunks passed to it before evaluating them. However, task
creation is expensive and its value is only realized upon a steal.
Steals are far less frequent than calls to \cd{forkjoin},
so our implementation ensures
that calling \cd{forkjoin} is cheap and expensive task creation is
deferred to the steal. In addition, one of the thunks is evaluated
immediately in the calling user-level thread, while only the other
thunk is exposed to other worker threads. This reduces the number of
user-level threads and thread switches by allowing a user-level thread
to evaluate a path of tasks.

In our implementation, we use a work-stealing scheduler that has been
annotated with heap management operations at the appropriate
places. Worker threads are implemented as OS-level pthreads and
user-level threads are implemented as the native user-level thread in
MLton. As the scheduler operates outside of the computation, the
objects it allocates, particularly in the \cd{schedule} function, do
not belong to any heap in the hierarchy.  Our implementation
has a separate ``global heap'' that is used to store scheduler
data.

\paragraph{Superheaps.}
The goal of making calls to \cd{forkjoin} cheap and deferring task
creation to steals directly informs our design and implementation of
hierarchical heaps.  Our implementation uses a structure
called a ``superheap'', which is associated with a user-level
thread. Superheaps contain a linked-list of heaps, each annotated with
its depth. This set of heaps corresponds to the heaps of the path of
tasks evaluated by the superheap's associated user-level thread.

On a call to \cd{forkjoin}, the current superheap's depth is
incremented and future allocations take place in the new heap created
for that depth. Once both thunks passed to \cd{forkjoin} are
completed, the depth is decremented. On depth decrement, the new heap
will be joined to its parent heap as per the algorithm. As the set of
heaps in the superheap is maintained as a linked list, and heaps are
added and joined in LIFO order, the depth increment and decrement
operations are very cheap.

If no steal occurs, both thunks passed to \cd{forkjoin} will be
evaluated in the newly created heap. If an idle worker thread steals
one of the thunks, it creates a new user-level thread and associated
superheap for that thunk. This superheap is then attached as a child
to the parent superheap in order for the runtime to be aware of the
complete hierarchy of heaps. When the stolen thunk is complete, it
will reactivate its parent task which can then merge the superheap and
extract the return value. Merging a superheap is just merging its set
of heaps, which is a simple linked-list operation.

\paragraph{Heaps.}
We implement a heap as a linked-list of variable-sized memory regions
called ``chunks''. This formulation enables efficient implementations
of key heap operations. Increasing heap size and joining heaps
(\cd{joinHeap}) are constant-time linked-list operations that do not
require objects to be copied. Finding the heap of an arbitrary pointer
(\cd{heapOf}) is implemented by looking up the chunk metadata using
address masking, which then contains a pointer to the heap associated
with that chunk.


\fi

\end{document}